\documentclass[aps,twocolumn,showpacs]{revtex4}
\usepackage{graphicx}

\begin{document}
\title{Influence of retardation effects on 2D magnetoplasmon spectrum}

\author{M. V. Cheremisin}
\affiliation{A.F.Ioffe Physical-Technical Institute,
St.Petersburg, Russia}
\date{\today}

\begin{abstract}
Within dissipationless limit the magnetic field dependence of
magnetoplasmon spectrum for unbounded 2DEG system found to
intersect the cyclotron resonance line, and, then approaches the
frequency given by light dispersion relation. Recent experiments
done for macroscopic disc-shape 2DEG systems confirm theory
expectations.
\end{abstract}


\maketitle

Plasma oscillations in two-dimensional electron gas(2DEG) were
first predicted in the middle 60th \cite{Stern}, and, then
observed experimentally in liquid helium system \cite{Grimes} and
silicon inversion layers \cite{Allen,Theis}. The recent pioneering
observation \cite{Kukushkin} of retardation effects in the
magnetoplasmon(MP) spectrum, discussed more than tree decades ago,
recommences the interest to the above problem. With retardation
effects accounted we analyze 2D MP spectrum first derived in
Ref.\cite{Chui}, and, then demonstrate the relevance of theory
predictions with respect to experimental data \cite{Kukushkin}.

\begin{figure}
\vspace*{0.5cm}
\includegraphics[scale=0.75]{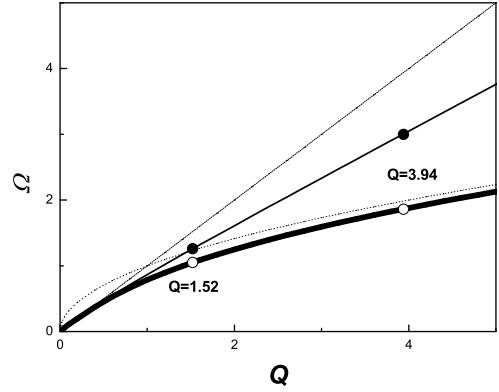}
\caption[]{\label{Fig1} Zero-field plasmon dispersion(bold line)
Asymptotes: light dispersion (dashed line), short-wavelength
plasmon limit (dotted line). Thin line represents the
$\Omega_{cr}(Q)$ when the condition $\omega=\omega_{c}$ is
satisfied. The calculated zero-field plasmon frequency(open dots)
and MP-CR intersection frequency(closed dots) are presented for
two different samples($Q=3.94$ and $Q=1.52$, see the text) used in
Ref\cite{Kukushkin}.}
\vspace*{-0.5cm}
\end{figure}

Let us assume unbounded 2D electron gas imbedded in a dielectric
in the presence of the perpendicular magnetic field. Following
Ref.\cite{Falko}, the Maxwell equations for in-plane components of
the electrodynamic potentials $\bf{A}, \phi$ yield
\begin{eqnarray}
\fbox{}\phi=4\pi\rho, \fbox{}{\bf A}=\frac{4\pi {\bf j}}{c},
\eqnum{5} \label{Maxwell} \\
\text{div}{\bf A}+\frac{\epsilon}{c}\frac{\partial \phi}{\partial t}=0, \nonumber \\
{\bf j}=-\sigma^{*}\left( \nabla \phi +\frac{1}{c}\frac{\partial
{\bf A}}{\partial t} \right), \nonumber
\end{eqnarray}
where $\fbox{}=\frac{\epsilon}{c^{2}}\frac{\partial^{2}}{\partial
t^{2}}-\Delta$ is the d'Lambert operator, $\sigma^{*}$ the
conductivity tensor. Assuming the magnetoplasmon $e^{(i{\bf
qr}-i\omega t)}$ propagated in 2DEG, and, then separating
longitudinal and transverse in-plane components of the vector
potential \cite{Falko}, 2D magnetoplasmon dispersion relation
yields:
\begin{equation}
\left( \frac{\epsilon}{2\pi}+\frac{i \sigma_{xx} \kappa}{\omega}
\right )
\left(\frac{1}{2\pi}-\frac{i \omega \sigma_{xx}}{c^{2}\kappa} \right )+ \frac{\sigma_{yx}^{2}}{c^{2}}=0, \\
\label{MP_dispersion}
\end{equation}
where $\kappa=\sqrt{q^{2}-\epsilon \frac{w^{2}}{c^{2}}}$. This
result is exactly that obtained by Chui \cite{Chui}. Within
dissipationless limit the components of the conductivity tensor
$\sigma_{xx}=\sigma_{yy}=\frac{i \omega
ne^{2}}{m(\omega^{2}-\omega^{2}_{c})}$,
$\sigma_{yx}=-\sigma_{xy}=\frac{i\omega_{c}}{\omega}\sigma_{xx}$
allow us to simplify Eq.(\ref{MP_dispersion}) as follows

\begin{equation}
(Q^{2}-\Omega^{2})^{\frac{1}{2}}=\sqrt{\frac{(1+\Omega^{2}_{c}-
\Omega^{2})^{2}}{4}+\Omega^{2}}-\frac{1+\Omega^{2}_{c}-\Omega^{2}}{2},
\label{MP_specrtum}
\end{equation}
where we introduce the dimensionless wave vector
$Q=\frac{qc}{\omega_{p}\sqrt{\epsilon}}$, frequency
$\Omega=\frac{\omega}{\omega_{p}}$ and cyclotron frequency
$\Omega_{c}=\frac{\omega_{c}}{\omega_{p}}$. Then,
$\omega_{p}=\frac{2\pi ne^{2}}{mc\sqrt{\epsilon}}$ is the plasma
frequency. In absence of the magnetic field Eq.(\ref{MP_specrtum})
reproduces the conventional \cite{Stern,Falko} zero-field
longitudinal plasmon dispersion relation $\epsilon=2\pi i
\sigma_{xx}\kappa/\omega$ as follows
$Q^{2}=\Omega^{2}+\Omega^{4}$. In the short-wavelength limit $Q
\gg 1$ one obtains square-root plasmon dispersion(dotted line in
Fig.\ref{Fig1}) $\omega=\sqrt{\frac{2\pi ne^{2}}{m \epsilon}q}$.
The opposite long-wavelength limit case $Q \ll 1$ corresponds to
light dispersion relation $\omega=\frac{cq}{\sqrt{\epsilon}}$
shown in Fig.\ref{Fig1} by the dashed line. Note, the authors
\cite{Kukushkin} demonstrate the excellent agreement between the
simplest zero-field 2D plasmon theory \cite{Stern,remark} and
experimental results. For different disc-geometry quantum well
samples the wave vector reported to relate to 2DEG disc radius via
$q=\frac{\alpha}{R}, \alpha=1.2$ in consistent with theory
$\alpha=3\pi/8$ \cite{Leavitt}.

\begin{figure}
\vspace*{0.5cm}
\includegraphics[scale=0.75]{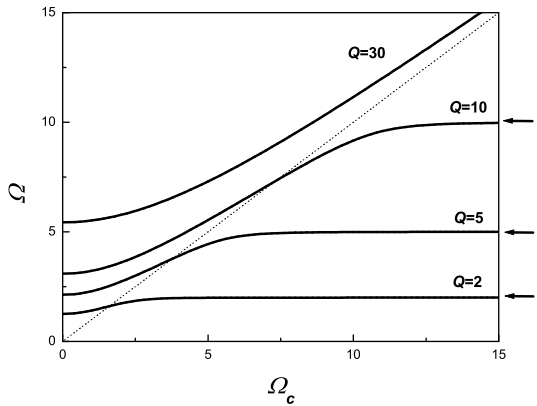}
\caption[]{\label{Fig2} 2D magnetoplasmon spectra vs dimensionless
cyclotron frequency $\Omega_{c}$ for different MP wave vector
$Q=2,5,10,30$. Arrows represent the light frequency
$\omega=\frac{cq}{\sqrt{\epsilon}}$. Cyclotron resonance is
represented by dotted line.} \vspace*{-0.5cm}
\end{figure}

The question we attempt to answer is whether retardation effects
should modify 2D magnetoplasmon spectrum. In actual fact,
Eq.(\ref{MP_specrtum}) demonstrate that at fixed $Q$ the plasma
frequency grows with magnetic field (see Fig.\ref{Fig2}), and,
then intersects the CR line. Note, this behavior is unexpected
within the edge magnetoplasmon formalism \cite{Fetter,Volkov}.
Substituting $\Omega=\Omega_{c}$ into Eq.(\ref{MP_specrtum}) we
derive the dependence(Fig.\ref{Fig1}, thin line) of the MP-CR
intersection frequency vs wave vector $\Omega_{cr}(Q)$. It to be
noted that low-field magnetoplasmon remains longitudinal at
$\omega_{c}<\omega$, and, then becomes transverse one when
$\omega_{c}>\omega$.

It is instructive to compare the lowest angular momentum MP
spectrum reported in \cite{Kukushkin} with that provided by the
present theory. For high-density GaAs/AlGaAs heterostructure
parameters ($n=2.54 \times 10^{11}$cm$^{-2}$, $m=0.067 m_{0},
R=0.5$mm, $\epsilon=12.8$) one obtains $\omega_{p}=51$GHz and
$Q=3.94$(see Fig.\ref{Fig1}). The calculated zero-field plasmon
frequency $f_{0}=\frac{\omega_{0}}{2\pi}=16$GHz and
$f_{cr}=\frac{\omega_{cr}}{2\pi}=24$GHz are well comparable with
experimental values $20$GHz and $32$GHz respectively. For higher
2D density sample($n=6.6 \times 10^{11}$cm$^{-2}$, $R=0.5$mm) we
obtain $\omega_{p}=132$GHz,$Q=1.52$, and then $f_{0}=22$GHz and
$f_{cr}=26$GHz compared to the respective experimental values
$27$GHz($36$GHz). In contrast, low-density sample($n=2.54 \times
10^{11}$cm$^{-2}, R=0.1$mm, $Q=109, f_{0}=14.6$GHz ) should
exhibit intersection of the MP spectrum with CR line at $B=2.52$T,
i.e. within magnetic field range, which is higher than that used
in Ref.\cite{Kukushkin}.

Further increase of the magnetic field results in saturation of
the magnetoplasmon spectrum(Fig.\ref{Fig2}) at certain frequency
given by the light dispersion relation
$\omega=cq/\sqrt{\epsilon}$. Experimentally, irrespective to 2DEG
density the larger disc-mesa samples($R=0.5$mm) demonstrate
\cite{Kukushkin} MP spectra(with the lowest radial and angular
momenta numbers) cut-off at certain frequency $f=50$GHz. The
latter coincides with frequency deduced from light dispersion
relation $f=\frac{1}{2\pi}\frac{qc}{\sqrt{\epsilon}}=51$GHz. Note,
in contrast to predicted MP spectrum saturation in strong fields,
the experiments \cite{Kukushkin} demonstrate intriguing zigzag
behavior which remains unexplained within our formalism.

In conclusion, we demonstrate the strong influence of retardation
effects on magnetoplasmon spectrum in unbounded 2DEG system. The
magnetic field dependence of MP spectrum found to intersect the
cyclotron resonance line, and, then approaches the frequency given
by light dispersion relation. The recent MP experiments in large
disc-shape 2DEG system confirm theory predictions.

This work was supported by RFBR(grant 03-02-17588) and
LSF(HPRI-CT-2001-00114, Weizmann Institute)


\begin{thebibliography}{99}
\bibitem{Stern} F. Stern, Phys. Rev. Lett. {\bf 18}, 546 (1967).

\bibitem{Grimes} C.C. Grimes and G.Adams, Phys. Rev. Lett. {\bf 36}, 145 (1976).

\bibitem{Allen} S.J. Allen, D.C.Tsui, R.A.Logan, Phys. Rev. Lett. {\bf 38}, 98 (1977).

\bibitem{Theis} T.N. Theis, J.P.Kotthaus, P.J.Stiles, Solid State Comm. {\bf 26}, 603 (1978).

\bibitem{Chui} K.W.Chui and J.J.Quinn, Phys. Rev. B {\bf 9}, 4724 (1974).

\bibitem{Kukushkin} I.V.Kukushkin $et al$, Phys. Rev. Lett. {\bf 90}, 156801 (2003).

\bibitem{Falko} V.I.Falko and D.E.Khmel'nitskii, Sov. Phys. JETP {\bf 68}, 1150 (1989).

\bibitem{Leavitt} R.P.Leavitt and J.W.Little, Phys.Rev.B {\bf 34}, 2450 (1986).

\bibitem{remark} In both 3D and 2D cases the edge plasmon frequency known to differ
with respect to bulk case(see K.W.Chui and J.J.Quinn, Phys.Rev.B
{\bf 5}, 4707 (1972) and \cite{Fetter,Volkov})

\bibitem{Fetter} A.L.Fetter, Phys.Rev.B {\bf 32}, 7676 (1985).

\bibitem{Volkov} V.A.Volkov and S.A.Mikhailov, Sov. Phys. JETP {\bf 67}, 1639 (1988).


\end{thebibliography}
\end{document}